\documentclass[10pt, draft]{article}
\usepackage{amsmath}
\usepackage{amssymb}
\usepackage{amsthm}
\usepackage{eucal}

\newtheorem{Theorem}{Theorem}

\newtheorem{Proposition}{Proposition}

\theoremstyle{remark}

\newtheorem*{Proof}{Proof}

\begin{document}
\title{More On Grover's Algorithm}
\author{Ken Loo\thanks{
   P.O. Box 9160, Portland, OR.\ 97207, look@sdf.lonestar.org} 
}
\date{}
\maketitle

\numberwithin{equation}{section}
\numberwithin{footnote}{section}
\numberwithin{Theorem}{section}
\numberwithin{Definition}{section}
\numberwithin{Proposition}{section}
\numberwithin{Corollary}{section}
\numberwithin{Remark}{section}
\numberwithin{Example}{section}

\begin{abstract}
   The goals of this paper are to show the following.
   First, Grover's algorithm
   can be viewed as a digital approximation to the analog quantum
   algorithm proposed in "An Analog Analogue of a Digital
   Quantum Computation", by E. Farhi and S. Gutmann, 
   Phys.\ Rev.\ A 57, 2403–-2406 (1998), quant-ph/9612026.
   We will call the above analog algorithm the Grover-Farhi-Gutmann
   or GFG algorithm.
   Second, the propagator of the GFG algorithm can be written
   as a sum-over-paths formula and given a sum-over-path interpretation,
   i.e., a Feynman path sum/integral.  We will use nonstandard analysis
   to do this.  Third, in the semi-classical limit $\hbar\to 0$,
   both the Grover and the GFG algorithms (viewed
   in the setting of the approximation in this paper) must run
   instantaneously.  Finally,
   we will end the paper with an open question.
   In "Semiclassical Shor's Algorithm", by P. Giorda, et al,
    Phys.\ Rev.\ A 70, 032303 (2004),
   quant-ph/0303037, the authors proposed building semi-classical
   quantum computers
   to run Shor's algorithm because the
   success probability of Shor's algorithm does not change
   much in the semi-classical limit.  We ask
   the open questions: In the semi-classical limit, does Shor's algorithm
   have to run instantaneously?
\end{abstract}

\section{\bf Introduction \label{Intro} } 
This paper attempts to answer the question:
what is a quantum algorithm?  In particular, what is
Grover's algorithm?  The views which we put forth in this
paper are that quantum mechanics is an analog, 
continuous-time theory, and Grover's algorithm is a digitization
of analog quantum mechanics.  In particular, 
it is a digitization of the analog quantum search
algorithm proposed by \cite{Far}.
The digitization is performed
in such a way so that the success probability of
the analog search algorithm is well preserved.  
We are not certain that this perspective of Grover's algorithm
is the correct one,  
but it is certainly true
that the theory of quantum mechanics describes 
quantum evolution in continuous time.    

The main goals of this
paper is to show some relationships between the Grover
algorithm and the analog quantum algorithm 
and elaborate more on
the analog algorithm.  
We will call the analog algorithm proposed by
\cite{Far} the GFG (Grover-Farhi-Gutmann) algorithm.
The three things which we will show in this paper are
the following.  We will show that Grover's 
algorithm can be viewed as a digital approximation to
the GFG algorithm.  
This interpretation comes about via
the Lie-Trotter product formula.  Second, We will
use the Lie-Trotter product formula and derive a
sum-over-paths formula for the propagator of
the GFG algorithm. We will then
give the sum-over-paths formula a Feynman path
summation (integral) interpretation.  This requires
some nonstandard analysis.  Finally, We will investigate
the semi-classical limit behaviors of the GFG
and the Grover algorithms
(in the digital approximation setting of this paper). 
This will lead us to ask an open question for
Shor's algorithm (for Shor's algorithm, 
see \cite{Nie} and references within).
We will assume that the reader is familiar with
both the Grover and the GFG algorithms, Feynman path integrals,
and a small amount of nonstandard analysis.   For the readers
who do not fall into the above category, 
we will lightly sketch the details when we go into
those subjects.

In a few sentences,
the GFG algorithm runs as follows.
Given a Hamiltonian $H_w = E|w\rangle\langle w|$, drive
the Hamiltonian $H_w$ into $H = H_w + H_s$
where 
\footnote{
In \cite{Far},
$H_s = E|s\rangle\langle s|$.  Hence, our evolution differs
from the one in \cite{Far} by a phase.}
$H_s = E\left(|s\rangle\langle s| - I\right)$.  
Evolve the initial state $|s\rangle$ via the evolution
$e^{-itH/\hbar}$ for some time $t = t^{'}$, then
$|\langle w |
e^{-it^{'}H/\hbar} |s\rangle |^2 \approx 1$,
i.e., the algorithm finds $|w\rangle$.

It turns out that it is possible to derive an 
interesting relationship
between the GFG algorithm just described and Grover's 
algorithm.  This relationship comes from the 
Lie-Trotter product formula and it naturally leads to the 
interpretation that Grover's algorithm is a digital approximation
of the GFG algorithm.
Using the Lie-Trotter product formula, we can
break the above evolution into
\begin{equation}
   \begin{split}
   \label{Trot-Intro}
   e^{-itH_s/\hbar} {}& = 
   \lim_{n\to\infty}
   \left(e^{-itH_s/n\hbar}
   e^{-itH_w/n\hbar} \right)^n \\
   {}&\approx
   \left(e^{-itH_s/k\hbar}
   e^{-itH_w/k\hbar} \right)^k, k >> 1 . 
   \end{split}
\end{equation}
By setting $t$ to the appropriate value, 
the second line of 
expression \ref{Trot-Intro} becomes a $k$-product of
the "Grover search engine"\footnote{This term was
coined by \cite{Chen}, see equation \ref{Engine}.}.  
In general, setting $t$ to the appropriate value 
destroys the Lie-Trotter approximation in \ref{Trot-Intro} in
the sense that the error is no longer bounded by inverse powers
of $k$.  The situation is remedied by 
setting $k$ to the proper value. 
Doing so, the success probability
of finding $|w\rangle$ in the GFG 
algorithm is well preserved (with error bounded
by inverse powers of $k$) in the approximation.
The analysis of the former error 
was previously done in \cite{Rol}.  We can attempt to 
derive a satisfying $\mathcal O\left(1\right)$
bound on the former error, see \cite{Rol} 
(we will briefly outline this
in section \ref{Ana}).  Doing this leads to the results
in \cite{Rol}, but this is not our goal.  
Our goal in this paper is to concentrate on
the analysis of the latter error.  
We wish to show that as the
size of the database to be search approaches 
infinity, the success probability of Grover's 
algorithm approaches the success probability
of the GFG algorithm, i.e., the error 
between the two probabilities is bounded by inverse powers of
the size of the database.
Because of this, we take on the view that Grover's algorithm 
is a digital approximation of the GFG algorithm.
In this sense, we can view Grover's
algorithm as a special case 
of the 
Lie-Trotter product cut-off
approximation for the GFG algorithm.
By special case, we mean that the
approximation is good for the
success probabilities with 
input vector $|s\rangle$ and output vector
$|w\rangle$, i.e., the approximation
is not global and it does not preserve
the wave functions.  
We will show this section \ref{Ana}.
Given the current state of the art,
whether the Grover algorithm is an approximation
of the GFG algorithm or the GFG algorithm is
an approximation of the Grover algorithm is 
a matter of taste.  We take on the former
interpretation because quantum mechanics
is an analog theory and the second line
of \ref{Trot-Intro} becomes Grover's algorithm
even though it only approximates the 
success probability of the GFG evolution.

Those who are familiar with Feynman path integral
techniques should recognize that this is an opportunity
for a sum-over-paths interpretation.  For more
on Feynman path integrals, see \cite{Alb, Cam1, Cam2, Fey, Fey2,
Klein, Sch}.
The traditional non-relativistic quantum mechanics
interpretation of the propagator $K\left(\vec x, \vec y, t\right)$
is that it is the probability amplitude of a particle
that starts at position $\vec x$ at time zero and ends up
at position $\vec y$ at time $t$.  The Feynman path integral
interpretation of the propagator is that for any classical
path that starts at $\vec x$ at time zero and ends at
$\vec y$ at time $t$, there is an amplitude associated
with that path and the total amplitude
$K\left(\vec x, \vec y, t\right)$
is the sum
of the amplitudes of all paths.  With physicists'
rigor, the Feynman path integral can be derived
by breaking up the evolution operator via the
Lie-Trotter product formula, interpreting
the limit as time slicing the physical evolution
and then going to the continuous time limit.
From a mathematician's point of view, this process
must be done in imaginary time since in real time,
there is no measure for which the Feynman path integral
converges (see \cite{Cam1, Cam2}).

Extending this into quantum computing, the propagator
$K\left(j,k,t\right)$ would be the probability
amplitude of starting at state $j$ at time zero
and ending up at state $k$ at time $t$.  A traditional
Feynman
path integral interpretation for this propagator
is somewhat tricky because the set of "positions"
for $j$ and $k$ is discrete and finite.  A
more serious problem for the propagator of the GFG 
algorithm
is that when time sliced, the summand contains the term
$\delta_{l_{m}, l_{m + 1}}$, which is 1 when the
state of the system is $|l_m\rangle$ at the $m^{th}$
time slice and is $|l_{m+1}\rangle$ at the
$m + 1^{st}$ time slice, and zero otherwise.  This
term is meaningless in the continuous time
limit.  It is for this reason that we use nonstandard
analysis to give the GFG propagator a sum-over-paths 
interpretation.  Using nonstandard analysis
on Feynman path integrals is not a new concept,
see \cite{Alb, Loo1, Loo2, Loo3, Loo4, Nak}.
For more on nonstandard analysis,
see \cite{Alb,Cut,Hur,Klein,Stroy}.  The amount of nonstandard analysis that
we will use is minimal.  We will use the nonstandard
reformulation of limits.  Given a sequence $r_n$ such that
\begin{equation}
  \label{lim}
  \lim_{n\to\infty} r_n = r , 
\end{equation}
the nonstandard equivalent of equation \ref{lim} is the following.
Given any infinite integer $\omega\in {}^*\mathbb N$,
where ${}^* \mathbb N$ is the nonstandard extension
of the natural numbers $\mathbb N$, $r_\omega$ is
infinitesimally close to $r$, and $st(r_\omega) = r$.
The operation $st$ known as the standard part.

Using nonstandard analysis, we will formulate
a sum-over-paths interpretation for the propagator
of the GHG algorithm as follows.  Fix an  
infinite $\omega\in{}^*\mathbb N$, time
slice $[0,t]$ into the set of discrete\footnote{Hyper-discrete
to be more precise} times
\begin{equation}
   T = \left\{0, t/\omega, 2t/\omega, \dots, t\right\}. 
\end{equation}
For each path\footnote{Internal path to be precise} 
$P:T\to\left\{1, 2, \dots, N\right\}$, where $N$ is the
size of the data base to be searched in Grover's algorithm,
we associate an amplitude to that path.  The propagator
is then infinitesimally close to
the sum\footnote{Internal sum to be precise} 
of the amplitudes of all such possible paths.
Further, this has to be true for all 
$\omega\in{}^*\mathbb N$.
We will show this in section \ref{Path}.

Finally, we will investigate the semi-classical behaviors
of the Grover algorithm and the GFG algorithm
in the setting of this paper.  We will show that
in the semi-classical limit, both algorithms must
run instantaneously.  We must point out that
if one day we figure out how to build a quantum
computer to run Grover's algorithm,
we do not know whether Grover's algorithm
will behave this way in the semi-classical limit.  
This is because in the setting of this
paper, we think of the Grover algorithm as a 
digital approximation
of the GFG algorithm, which may or may not
be the appropriate point of view.  Recently, the authors in
\cite{Semi} proposed building semi-classical quantum computers
for Shor's
algorithm.  They showed
that in the semi-classical limit, the success
probability of Shor's algorithm does not change much.
At this point, we ask an open question.  
In the semi-classical limit,
does Shor's algorithm have to run instantaneously?

\section{\bf The Grover Algorithm \label{Gate}}
In this section, we will outline Grover's 
algorithm.  For more details on the algorithm, 
see \cite{Chen,Far, Grov, Nie}.  The material
in this section is taken from \cite{Chen}.
We are given a function
$f:\left\{0,\dots,N - 1\right\}\to\left\{0,1\right\}$
such that there exist a $w$ with $f(w) = 1$ and
$f(a) = 0$ for $a \neq w$.  The goal is to use a quantum
computer to find $w$.

Grover's algorithm is as follows.  Let $U_f$ be the unitary
operator defined by
\begin{equation}
    U_f|a\rangle = \left(-1\right)^{f(a)}|a\rangle =
    I - 2P_w,\,\, P_w = |w\rangle\langle w|,   
\end{equation}
and let
\begin{equation}  
  |s\rangle = \dfrac{1}{\sqrt{N}}\sum_{a = 0}^{N - 1} |a\rangle .
\end{equation}
Define $U_s$ by
   \footnote{In \cite{Chen}, 
    $U_s$ is defined as $U_s = I - 2|s\rangle\langle s|$.}. 
\begin{equation}
   U_s = 2|s\rangle\langle s| - I = 2P_s - I,\,\, P_s
   = |s\rangle\langle s|, 
\end{equation}
and let 
\begin{equation}
    \label{Engine}
    U_G = U_sU_f
\end{equation}
be the "Grover search engine"\footnote{The term was coined by
\cite{Chen}}.
Let 
\begin{equation}
   |r\rangle = \dfrac{1}{\sqrt{1 - (1/N)}}\left(|s\rangle
    - \dfrac{|w\rangle}{\sqrt{N}}\right),
\end{equation}
and let $\mathcal V$ be the subspace spanned by
$\left\{|w\rangle, |r\rangle\right\}$.  

\begin{Theorem}
\label{Grov-Thm1}
The subspace
$\mathcal V$ is an invariant two-dimensional subspace 
of $U_G$.  Further, with respect to 
the orthonormal basis 
$\left\{|w\rangle, |r\rangle\right\}$ in the invariant
subspace $\mathcal V$, $U_G$ admits the unitary
matrix representation
\begin{equation}
  U = \left[
        \begin{matrix}
         \frac{N - 2}{N} & -\frac{2\sqrt{N - 1}}{N} \\
         \frac{2\sqrt{N - 1}}{N} &
         \frac{N - 2}{N} 
         \end{matrix}
      \right] = 
       \left[
        \begin{matrix}
         \cos{\theta} & -\sin{\theta} \\
         \sin{\theta} &
         \cos{\theta} 
         \end{matrix}
      \right], \quad \theta = 
      \sin^{-1}{\left( 
         \frac{2\sqrt{N - 1}}{N} 
       \right) } .  
\end{equation}
\end{Theorem}
\begin{Proof}
  See \cite{Chen}. \qed
\end{Proof}

\begin{Theorem} 
\label{Grov-Thm2}
Let $U_G, |s\rangle, \dots$ be as previously
defined, then
\begin{equation}
      U^k|s\rangle =  
       \left[
        \begin{matrix}
         \cos{\theta} & -\sin{\theta} \\
         \sin{\theta} &
         \cos{\theta} 
         \end{matrix}
      \right]^k \left(
      \begin{matrix}
              \frac{1}{\sqrt{N}} \\
             \sqrt{\frac{N - 1}{N}}
      \end{matrix}\right) = 
      \left(
      \begin{matrix}
              \cos{\left(k\theta + \alpha\right)} \\
              \sin{\left(k\theta + \alpha\right)} 
      \end{matrix}\right),\quad \alpha = 
      \cos^{-1}{\left(\frac{1}{\sqrt{N}}\right)}.  
\end{equation} 
The probability of reaching the state
$|w\rangle$ after $k$ iterations is 
\begin{equation}
   \label{Grov-Prob}
   \mathcal P_k = |\langle w| U^k |s\rangle|^2 =
   \cos^2{\left(k\theta + \alpha\right)}.
\end{equation}
Further, at 
$k = \frac{\pi\sqrt{N}}{4}, \mathcal P_k\approx 1$.
\end{Theorem}  
\begin{Proof}
See \cite{Chen}. \qed
\end{Proof}

Basically, theorems \ref{Grov-Thm1} and
\ref{Grov-Thm2} say that the
"Grover search engine" rotates the vector  
$|s\rangle$ in the subspace $\mathcal V$
to the vector $|w\rangle$ after $k$ iterations.

\section{\bf The GFG Algorithm \label{GFG}}
In this section, we will outline the GFG algorithm.
For more details on the materials in this section, 
see \cite{Chen, Far}.  The GFG algorithm is similar to
the Grover algorithm in the sense that they are both 
search algorithms.  
As stated in \cite{Far},
the GHG algorithm solves the following problem.
Given a Hamiltonian $H_w = E|w\rangle\langle w|$
where $|w\rangle$ is a basis vector (the
Hilbert space has dimension $N$), find $|w\rangle$.

The GFG algorithm solves the above problem as follows.
Drive the Hamiltonian $H_w$ into the Hamiltonian 
   \footnote{For our purpose, we will want 
    $H = H_w + H_s \equiv H_w + \bar H_s - EI$. 
    Hence the notation $\bar H$ and $\bar H_s$.}, 
\begin{equation}
   \bar H = H_w + \bar H_s ,\,\, 
\end{equation}
where
\begin{equation}
  \bar H_s = E|s\rangle \langle s|, \quad 
  |s\rangle = \dfrac{1}{\sqrt{N}}\sum_{a = 0}^{N - 1} |a\rangle .
\end{equation}
Let
\begin{equation}
   |r\rangle = \dfrac{1}{\sqrt{1 - (1/N)}}\left(|s\rangle
    - \dfrac{|w\rangle}{\sqrt{N}}\right),
\end{equation}
then, as in the Grover algorithm,
the two dimensional subspace $\mathcal V$ spanned by
the orthonormal basis $\left\{|r\rangle, |w\rangle\right\}$ 
is an invariant
subspace of the evolution under $\bar H$.  
In other words\nolinebreak
\footnote{In the quantum computing literature, the value of $\hbar$
is usually taken to be 1.  Because we will be taking semi-classical
limits, we will not take $\hbar = 1$.  See section \ref{Semi}.},
\begin{equation}
  e^{-it\bar H/\hbar}\mathcal V = \mathcal V.
\end{equation}
In the subspace $\mathcal V$ with respect to this basis, 
the evolution admits the matrix
representation
\begin{equation}
   e^{-it\bar H/\hbar} =
   e^{-\frac{iEt}{\hbar}}\left[\begin{matrix}
   \cos{\frac{Eyt}{\hbar}} - iy\sin{\frac{Eyt}{\hbar}} & 
     -\sqrt{1 - y^2}i\sin{\frac{Eyt}{\hbar}} \\
   -\sqrt{1 - y^2}i\sin{\frac{Eyt}{\hbar}} & 
     \cos{\frac{Eyt}{\hbar}} + iy\sin{\frac{Eyt}{\hbar}} 
   \end{matrix} 
   \right], y = \sqrt{\frac{1}{N}}.
\end{equation} 
For $t > 0$ and with the initial state $|s\rangle$,
the system evolves as 
\begin{equation}
  \begin{split}
  \phi(t) {}& = e^{-it\bar H/\hbar}|s\rangle \\
   {}& =
   e^{-\frac{iEt}{\hbar}}\left\{
   \left[y\cos{\frac{Eyt}{\hbar}} - i\sin{\frac{Eyt}{\hbar}}
    \right]|w\rangle
   + \sqrt{1 - y^2}\cos{\frac{Eyt}{\hbar}}|r\rangle\right\}.
   \end{split}
\end{equation}
The probability of measuring $|w\rangle$ is then
\begin{equation}
   \mathcal P(t) = 
  |\langle w | \phi(t)|^2 = 
  |\langle w| e^{-it\bar H/\hbar}|s\rangle|^2 =
   \sin^2{\frac{Eyt}{\hbar}} +
   y^2\cos^2{\frac{Eyt}{\hbar}},\,\,  y = \sqrt{\frac{1}{N}} .
\end{equation}
Hence, running the system for time 
$t = \frac{\pi\hbar\sqrt{N}}{2E}$    
will measure $|w\rangle$ with probability 1.

For our purposes, we want to work with the Hamiltonian
\begin{equation}
H = H_w + \bar H_s - EI \equiv H_w + H_s
= E\left(|w\rangle\langle w| + |s\rangle\langle s| - I\right)
= E\left(P_w + P_s - I\right)
\end{equation}
The evolution of the Hamiltonian $H$ differs from
the evolution of $\bar H$ by a factor of 
$e^{iEt/\hbar}$.
\begin{Theorem} 
\label{GFG-Thm1}
Let
\begin{equation}
H = H_w + \bar H_s - EI \equiv H_w + H_s
= E\left(|w\rangle\langle w| + |s\rangle\langle s| - I\right)
= E\left(P_w + P_s - I\right) .
\end{equation}
The space $\mathcal V$ is an
invariant subspace of the evolution $e^{-itH/\hbar}$.
In the orthonormal basis $\left\{|w\rangle, |r\rangle\right\}$,
$e^{-itH/\hbar}$ admits the unitary matrix representation
\begin{equation}
   e^{-itH/\hbar} =
   \left[\begin{matrix}
   \cos{\frac{Eyt}{\hbar}} - iy\sin{\frac{Eyt}{\hbar}} &
     -\sqrt{1 - y^2}i\sin{\frac{Eyt}{\hbar}} \\
   -\sqrt{1 - y^2}i\sin{\frac{Eyt}{\hbar}} &
     \cos{\frac{Eyt}{\hbar}} + iy\sin{\frac{Eyt}{\hbar}}
   \end{matrix}
   \right], y = \sqrt{\frac{1}{N}}.
\end{equation}
For $t > 0$ and with the initial state $|s\rangle$,
the system evolves as
\begin{equation}
  \begin{split}
  \phi(t) {}& = e^{-itH/\hbar}|s\rangle \\
   {}& =
   \left\{
   \left[y\cos{\frac{Eyt}{\hbar}} - i\sin{\frac{Eyt}{\hbar}}
    \right]|w\rangle
   + \sqrt{1 - y^2}\cos{\frac{Eyt}{\hbar}}|r\rangle\right\}.
   \end{split}
\end{equation}
The probability of measuring $|w\rangle$ is
\begin{equation}
   \label{GFG-Prob}
   \mathcal P(t) =
  |\langle w | \phi(t)|^2 =
  |\langle w| e^{-itH/\hbar}|s\rangle|^2 =
   \sin^2{\frac{Eyt}{\hbar}} +
   y^2\cos^2{\frac{Eyt}{\hbar}},\,\,  y = \sqrt{\frac{1}{N}} .
\end{equation}
Hence, running the system for time
$t = \frac{\pi\hbar\sqrt{N}}{2E}$
will measure $|w\rangle$ with probability 1.
\end{Theorem}
\begin{Proof} The evolution of the Hamiltonian $H$ is
$e^{-itH/\hbar} = e^{iEt/\hbar}e^{-it\bar H/\hbar}$. \qed
\end{Proof}

\section{\bf Digitizing The GFG Algorithm\label{Ana}}
In this section, we will digitize the GFG algorithm.
The philosophy which we adapt here is that we are aware
of the GFG algorithm and have the Lie-Trotter product formula
at our disposal.
We will see that under this philosophy, digitizing
the GFG algorithm naturally leads to the Grover algorithm.
For those who are familiar with Feynman path integral methods,
it is also possible to view the Grover algorithm as a by product
of giving the GFG algorithm a Feynman path integral interpretation.
Of course this point of view is hindsight since the development
of the two algorithms came in the reverse order and we are aware
of Grover's algorithm.  It is not surprising that this
is hindsight since Feynman path integral methods are in
general more complex than operator methods
\footnote{For example, it was many years after
Feynman's formulation of the Feynman path integral
before the hydrogen atom path integral was solved (with
physicist's rigor).  Feynman himself was not able to
solve the hydrogen atom path integral.  
See \cite{Klein} and references within.}. 
We take on the approach in this section
because historically, quantum mechanics is a time dependent 
analog theory and because it is pedagogical in answering the question:
what is a quantum algorithm?

We start with the Lie-Trotter produce formula.  For our purposes,
we only need the finite dimensional version, which is the Lie
product formula, see \cite{Reed}.
\begin{Theorem}
  \label{Trot}
 Let $A$ and $B$ be self-adjoint, finite-dimensional matrices, then
 \begin{equation}
   e^{-i\eta\left(A + B\right)} =
   \lim_{n\to\infty} \left[e^{-i\eta A/n} e^{-i\eta B/n}\right]^n .
  \end{equation}
Furthermore, for any $k$,
\begin{equation}
  ||e^{-i\eta\left(A + B\right)} - [e^{-i\eta A/k}e^{-i\eta B/k}]^k|| 
   \leq \dfrac{\eta^2 \mathcal O\left(||[A,B]||\right)}{k},  
\end{equation}
where $[A,B] = AB - BA$.
\end{Theorem}
\begin{Proof} Let $S_k = e^{-i\eta\left(A + B\right)/k}$, and
  $T_k = e^{-i\eta A/k}e^{-i\eta B/k}$.  Then 
\begin{equation}
   \label{equ4.1}
   \begin{split}
   ||S^k_k - T^k_k|| {}& = \bigg|\bigg|\sum_{m = 0}^{k - 1}
   S^m_k\left(S_k - T_k\right)T^{k - 1 - m}_k\bigg|\bigg| \\
   {}& \leq k||S^m_k|| ||T^{k - 1 - m}_k||
   ||\left(S_k - T_k\right)|| = 
   k||\left(S_k - T_k\right)|| . 
   \end{split}
\end{equation}
Further, using the Campbell-Baker-Hausdorff approximation yields
\begin{equation}
   \label{equ4.2}
   ||\left(S_k - T_k\right)|| = 
   \frac{\eta^2 \mathcal O\left(||[A, B]||\right)}{k^2} .
\end{equation}
Expressions \ref{equ4.1} and \ref{equ4.2} imply
\begin{equation}
  ||e^{-i\eta\left(A + B\right)} - 
    [e^{-i\eta A/k}e^{-i\eta B/k}]^k|| =
   ||S^k_k - T^k_k|| \leq
   \frac{\eta^2 \mathcal O\left(||[A,B]||\right)}{k} \to 0. \qed
\end{equation}    
\end{Proof}
Theorem \ref{Trot} is the first step towards a Feynman
path integral interpretation for the GFG algorithm.  
We will derive a sum-over-paths interpretation in
section \ref{Path}.  We continue with our attempt
to digitize the GHG algorithm.

\begin{Proposition} 
\label{Prop1}
Let $P$ be a projection operator,
then
\begin{equation}
  e^{-i\pi P} = I - 2P 
\end{equation}
\end{Proposition}
\begin{Proof}
Power expanding and using the fact that $P$ is
a projection, we get
\begin{equation}
  e^{-i\pi P} = \sum_{n = 0}^\infty \dfrac{\left(-i\pi P\right)^n}{n!}
  = I + \sum_{n = 1}^\infty \dfrac{\left(-i\pi\right)^n}{n!} P
  = I + (e^{-i\pi} - 1)P = I - 2P. \qed
\end{equation}
\end{Proof}
\begin{Proposition} Let $P_w = |w\rangle\langle w|$,
   \label{Prop2}
then $U_f = e^{-i\pi P_w}$.
\end{Proposition}
\begin{Proof} This is an application of proposition \ref{Prop1} by
writing $U_f = I - 2P_w$. \qed
\end{Proof}
\begin{Proposition} Let $P_s = |s\rangle\langle s|$,
   \label{Prop3}
then $U_s = e^{-i\pi\left( P_s - I\right)}$.
\end{Proposition}
\begin{Proof} This is an application of proposition \ref{Prop1} by
writing $U_s = -(I - 2P_s) = e^{i\pi I}e^{-i\pi P_s}
=e^{-i\pi\left( P_s - I\right)}$.
\qed
\end{Proof}

We are now ready to digitize the GHG algorithm presented
in theorem \ref{GFG-Thm1}.  The Hamiltonian which we use
for the GFG algorithm is 
$H = E(P_s - I +  P_w)$,
which produces the evolution
$e^{-itE\left(P_s - I + P_w\right)/\hbar}$.  Using
the Lie-Trotter product formula on this evolution yields
\begin{equation}
   e^{-itE\left(P_s - I + P_w\right)/\hbar} = 
   \lim_{n\to\infty}\left(e^{-itE\left(P_s - I\right)/n\hbar}
   e^{-itEP_w/n\hbar}\right)^n .
\end{equation}
Further, for large $k$, we can approximate the evolution via
\begin{equation}
   \label{Trot-Appr}
   e^{-itE\left(P_s - I + P_w\right)/\hbar} \approx 
   \left(e^{-itE\left(P_s - I\right)/k\hbar}
   e^{-itEP_w/k\hbar}\right)^k ,
\end{equation}
where the error in the approximation is of order
$\frac{t^2E^2}{\hbar^2 k}$.  For $k >> \frac{t^2E^2}{\hbar^2}$,
this approximation works well.  Suppose we let 
$\frac{tE}{\hbar} = k\pi$, then according to 
propositions \ref{Prop2} and \ref{Prop3},
he right hand side
of approximation \ref{Trot-Appr} becomes 
$\left(U_sU_f\right)^k$, which is a $k$-product
of the "Grover search engine".  
With $\frac{tE}{\hbar} = k\pi$, 
the error bound in approximation \ref{Trot-Appr} is 
no longer in inverse powers of $k$, 
it is $\mathcal O\left(1\right)$.  This can be seen as follows.
The Lie-Trotter product formula tells us that the error
is bounded by
\begin{equation}
   \label{error}
  \frac{t^2\mathcal O\left(||[P_s - I, P_w]||\right)}{k} . 
\end{equation}
Further, 
\begin{equation}
  ||[P_s - I, P_w]|| = \mathcal O\left(\dfrac{1}{\sqrt{N}}\right),
\end{equation}
and for the GFG and the Grover algorithms,
\begin{equation}
  t = \mathcal O\left(\sqrt{N}\right),
  k = \mathcal O\left(\sqrt{N}\right).
\end{equation}
Hence, \ref{error} becomes $\mathcal O\left(1\right)$.  This 
result, which is not our goal, was obtained 
by \cite{Rol}.  Our
goal here is to obtain error bounds in inverse powers of $k$.
Thus, in the asymptotic limit on the size of the database to be searched,
the success probability of the Grover algorithm approaches 
the success probability of the GFG algorithm.

As far as running the GFG 
algorithm is concerned,
we are interested in the probability
\begin{equation}
   \label{GFG-Prob1}
   \mathcal P(t) =
  |\langle w| e^{-it H/\hbar}|s\rangle|^2 =
   \sin^2{\frac{Eyt}{\hbar}} +
   y^2\cos^2{\frac{Eyt}{\hbar}},\,\,  y = \sqrt{\frac{1}{N}} .
\end{equation}
Thus, we settle for 
values of $t$'s and $k$'s such that the following
three properties are satisfied.  First,  
$t = \mathcal O\left(\sqrt{N}\right)$.
Second, the right
hand side of \ref{Trot-Appr} becomes a $k$-product
of the "Grover search engine".  Third,
with those values of $t$'s and $k$'s, the probabilities
given by equations \ref{GFG-Prob1} and 
\ref{Grov-Prob} are close to each other with error
bounded by inverse powers of $k$. 
In other words, rather than globally
digitize the GFG evolution operator, 
we settle for 
digitizing the operator in such a way so that
the success probability of the GFG algorithm
is well preserved.  More simply, we wish to find
correct combinations of $t$'s and $k$'s so that
\begin{equation}
   \begin{split}
   1 {}& \approx
   \mathcal P(t) =
  |\langle w| e^{-it H/\hbar}|s\rangle|^2 =
  \bigg|\langle w\big|
   \lim_{n\to\infty}
   \left(e^{-itE\left(P_s - I\right)/n\hbar}
   e^{-itEP_w/k\hbar}\right)^n
   \big|s\rangle\bigg|^2 \\
   {}& \approx
  \bigg|\langle w\big|
  \left(e^{-itE\left(P_s - I\right)/k\hbar}
   e^{-itEP_w/k\hbar}\right)^k
   \big|s\rangle\bigg|^2 =
  \big|\langle w\big|
   \left(U_sU_f\right)^k
   \big|s\rangle\big|^2 ,
  \end{split}
\end{equation}
where the error of the second approximation 
is bounded by inverse powers
of $k$.
\begin{Theorem}
\label{Major-Thm}
Let $l, m\in\mathbb N$ be such that
$l, m << \sqrt{N}$, $m$ is odd,  and
\begin{equation}  
   \frac{4\pi l}{\pi - 2} = m + \epsilon ,
\end{equation}
where $|\epsilon| \approx 0$. 
Let $k = \left[\frac{2\pi l\sqrt{N}}{\pi - 2}\right]$
be the nearest integer to 
$\frac{2\pi l\sqrt{N}}{\pi - 2}$, and let
$t = \frac{k\pi\hbar}{E}$.  
Finally, let 
\begin{equation}
  \delta = \max{\left\{\epsilon, \sqrt{\frac{1}{N}}\right\}}.
\end{equation}
Then, 
\begin{equation}
   \label{Proof1}
   \left(e^{-itE\left(P_s - I\right)/k\hbar}
   e^{-itEP_w/k\hbar}\right)^k = 
   \left(U_sU_f\right)^k, 
\end{equation}
\begin{equation}
   \label{Proof2}
   \begin{split}
   \mathcal P(t) {}& =
  |\langle w| e^{-it H/\hbar}|s\rangle|^2 \\
   {}& = \sin^2{\left(\frac{Et}{\sqrt{N}\hbar}\right)} +
   \frac{1}{N}\cos^2{\left(\frac{Et}{\sqrt{N}\hbar}\right)} = 1 + 
    \mathcal O\left(\delta^2\right),
  \end{split}
\end{equation}
and
\begin{equation}
   \label{Proof3}
   \begin{split}
   \mathcal P_k {}& = 
   \bigg|\big\langle w \big| \left(e^{-itE\left(P_s - I\right)/k\hbar}
   e^{-itEP_w/k\hbar}\right)^k \big|s\big\rangle \bigg|^2 \\
    {}& = \big|\langle w|\left(U_sU_f\right)^k|s\rangle \big|^2 =  
      \cos^2{\left(k\theta + \alpha\right)} = 
   \mathcal P(t) + \mathcal O\left(\frac{1}{N}\right)
   \end{split}
\end{equation}
\end{Theorem}
\begin{Proof} Equation \ref{Proof1} follows from
propositions \ref{Prop2} and \ref{Prop3}.  
Recall that (see \cite{Chen})
\begin{equation}
      \alpha =
      \cos^{-1}{\left(\frac{1}{\sqrt{N}}\right)} =
      \frac{\pi}{2} + \mathcal O\left(\sqrt{\frac{1}{N}}\right),
\end{equation}
\begin{equation}
  \theta =
      \sin^{-1}{\left(
         \frac{2\sqrt{N - 1}}{N}
       \right) } = 2\sqrt{\frac{1}{N}} + 
       \mathcal O\left(\frac{1}{N^{3/2}}\right) .
\end{equation}
With the above values of $k$, we have
\begin{equation}
     k\theta = \frac{2\pi l\sqrt{N}}{\pi - 2}2\sqrt{\frac{1}{N}} 
     + \mathcal O\left(\sqrt{\frac{1}{N}}\right) =
     \frac{4\pi l}{\pi - 2} 
     + \mathcal O\left(\sqrt{\frac{1}{N}}\right) .
\end{equation}
Further,
\begin{equation}
   \frac{Et}{\sqrt{N}\hbar} = 
   \frac{k\pi}{\sqrt{N}} =
      \frac{2\pi l\sqrt{N}}{\pi - 2}
      \frac{\pi}{\sqrt{N}}
     + \mathcal O\left(\sqrt{\frac{1}{N}}\right) =
     \frac{2\pi^2 l}{\pi - 2} 
     + \mathcal O\left(\sqrt{\frac{1}{N}}\right) .
\end{equation}
Then,
\begin{equation}
  k\theta - \frac{Et}{\sqrt{N}\hbar} = 
  \frac{4\pi l}{\pi - 2} -
     \frac{2\pi^2 l}{\pi - 2} 
     + \mathcal O\left(\sqrt{\frac{1}{N}}\right) =
     -2\pi l
     + \mathcal O\left(\sqrt{\frac{1}{N}}\right) .
\end{equation}
Hence,
\begin{equation}
  \begin{split}
  \mathcal P_k {}& =
   \bigg|\big\langle w \big| \left(e^{-itE\left(P_s - I\right)/k\hbar}
   e^{-itEP_w/k\hbar}\right)^k \big|s\big\rangle \bigg|^2 \\
    {}& = \big|\langle w|\left(U_sU_f\right)^k|s\rangle \big|^2 =
      \cos^2{\left(k\theta + \alpha\right)}  \\
    {}& = \cos^2{\left[\frac{Et}{\sqrt{N}\hbar} - 2\pi l +
        \frac{\pi}{2} 
     + \mathcal O\left(\sqrt{\frac{1}{N}}\right) \right]} \\
    {}& = \sin^2{\left(\frac{Et}{\sqrt{N}\hbar}\right)} 
     + \mathcal O\left(\frac{1}{N}\right) \\
    {}& = \mathcal P(t)
     + \mathcal O\left(\frac{1}{N}\right) .
   \end{split}
\end{equation}
As for $P(t)$, we have
\begin{equation}
  \label{P-eqn}
  \begin{split}
  P(t) {}& = 
  \sin^2{\left(\frac{Et}{\sqrt{N}\hbar}\right)} +
   \frac{1}{N}\cos^2{\left(\frac{Et}{\sqrt{N}\hbar}\right)} =
  \sin^2{\left(\frac{Et}{\sqrt{N}\hbar}\right)} 
     + \mathcal O\left(\frac{1}{N}\right) \\
   {}& = \sin^2{\left(\frac{2\pi^2 l}{\pi - 2} +
     \mathcal O\left(\sqrt{\frac{1}{N}}\right) 
      \right)} 
     + \mathcal O\left(\frac{1}{N}\right) \\
     {}&= \sin^2{\left(\frac{\pi}{2}\frac{4\pi l}{\pi - 2} +
     \mathcal O\left(\sqrt{\frac{1}{N}}\right) \right)} 
     + \mathcal O\left(\frac{1}{N}\right) \\
     {}&= \sin^2{\left(\frac{\pi}{2}m +
     \mathcal O\left(\delta\right) \right)} 
     + \mathcal O\left(\frac{1}{N}\right). \\
   \end{split}
\end{equation}
since $m$ is odd, \ref{P-eqn} becomes 
\begin{equation}
   P(t) = 1 + \mathcal O(\delta^2).  \qed
\end{equation}
\end{Proof}

In light of theorem \ref{Major-Thm}, 
we can interpret Grover's algorithm
as a digitization of the GFG algorithm
via the Lie product formula.  While
the wave function is not preserved under
the digitization, the success probability is
well preserved.  We can also view the "Grover
search engine" as 
$e^{-itE\left(P_s - I\right)/k\hbar}e^{-itEP_w/k\hbar}$
with $t$ and $k$ as specified in theorem \ref{Major-Thm}.
Obviously, this view might only be useful for theoretical analysis
since it would not be efficient to run Grover's algorithm
by running $t = \mathcal O\left(\sqrt{N}\right)$ and then
$k = \mathcal O\left(\sqrt{N}\right)$ copies of the "Grover
search engine".

We end this section with an numerical example of 
theorem \ref{Major-Thm}.  Suppose $N = 2^{30}$, and
$l = 1$.  Then,
\begin{equation}
   \label{Values}
   \begin{split}
   {}& \frac{4\pi l}{\pi - 2} \approx 11 + .00775358 
   \equiv m + \epsilon, \\
   {}&\delta = \max{\left\{\epsilon, \frac{1}{\sqrt{N}}\right\}}
             = \max{\left\{.00775358, \frac{1}{2^{15}}\right\}} = 
               \epsilon, \\
   {}& \alpha = \cos^{-1}{\left(\frac{1}{2^{15}}\right)} 
       \approx 1.57076581, \\
   {}& \theta = 
        \sin^{-1}{\left(\frac{2\sqrt{2^{30} - 1}}{2^{30}}\right)} 
       \approx .00006104, \\
   {}& k = \left[\frac{2\pi 2^{15}}{\pi - 2}\right] = 
          180351, \\
   {}&\frac{Et}{\sqrt{N}\hbar} = \frac{k\pi}{\sqrt{N}} 
      \approx 17.29093889 .
   \end{split}
\end{equation}
With the values given in \ref{Values},
we have
\begin{equation}
   \begin{split}
   P(t) {}& \approx \sin^2{\left(17.29093889\right)} +
     \frac{1}{2^{30}}
   \cos^2{\left(17.29093889\right)} \\
    {}& \approx .99985167
   = 1 + \mathcal O\left(\epsilon^2\right)
   = 1 + \mathcal O\left(.00006012\right) .
  \end{split}
\end{equation}
Lastly,
\begin{equation}
   \begin{split}
   P_k {}& = \cos^2{\left(k\theta + \alpha\right)} \approx
   \cos^2{\left(180351*.00006104 + 1.57076581\right)} \\
   {}& \approx .99983048 = 
    .99985167 + \mathcal O\left(.00003052\right) =
   P(t) + \mathcal O\left(\frac{1}{N}\right) .  
   \end{split}
\end{equation}
Notice that in the example, we did not specify the
value of $E$.  Hence, we do not have a numerical
value for $t$.

\section{\bf Sum-Over-Paths Interpretation\label{Path}}
In this section, we will derive a sum-over-paths (Feynman
path integral) interpretation for the GFG algorithm.
Recently, sum-over-paths techniques were applied to
quantum algorithms, see \cite{Daw}.  
The sum-over-paths interpretation
of the work in \cite{Daw} is somewhat unconventional compared to
traditional Feynman path integral methods 
(see \cite{Fey, Fey2, Klein, Sch}) in
the sense that the number of paths in \cite{Daw} are finite.  
It is the goal
of this section to derive a sum-over-paths interpretation
for the GFG algorithm using more traditional methods.
In traditional Feynman path integrals, breaking up
the evolution operator $e^{-it(H_1 + H_2)}$ via the
Lie-Trotter product formula is the first step towards
the derivation of the Feynman path integral.  

We are interested in deriving a sum-over-paths formula for
the propagator of the GFG algorithm.  For any
quantum states $|x\rangle$ and $|y\rangle$,
we can write
\begin{equation}
  \langle x| e^{-itH/\hbar} |y \rangle  =
  \sum_{j,k} \langle x|j\rangle\langle j |
  e^{-itH/\hbar} |k \rangle \langle k |y \rangle . 
\end{equation}
The propagator $K(j,k,t)$ is defined by
\begin{equation}
   \label{Propa}
   K(j,k,t) \equiv
  \langle j |
  e^{-itH/\hbar} |k \rangle . 
\end{equation}

The traditional non-relativistic quantum mechanics
interpretation of the propagator $K\left(\vec x, \vec y, t\right)$
is that it is the probability amplitude of a particle
that starts at position $\vec x$ at time zero and ends up
at position $\vec y$ at time $t$.  The Feynman path integral
interpretation of the propagator is that for any classical
path from $\vec x$ to $\vec y$ at time zero to time $t$,
an amplitude is assigned to that path and the total amplitude
$K\left(\vec x, \vec y, t\right)$ is the sum of all such
path amplitudes.  We are interested in obtaining a similar
interpretation for the propagator in equation \ref{Propa}. 

\begin{Proposition} \label{Path-Prop1}
Let $|l\rangle$
be a basis vector of the computational basis, then
\begin{equation}
   e^{\frac{-itE|w\rangle\langle w|}{n\hbar}}
   | l \rangle = e^{\frac{-itEf(l)}{n\hbar}}|l\rangle . 
\end{equation}
\end{Proposition}
\begin{Proof} We have
\begin{equation}  
   \begin{split}
   e^{\frac{-itE|w\rangle\langle w|}{n\hbar}}
    | l\rangle {}& =
    | l\rangle +
  \sum_{r = 1}^\infty \left(\dfrac{-itE}{n\hbar}\right)^r
  \dfrac{1}{r!}
   P_w |l\rangle =
    | l\rangle + \left(e^{\frac{-itE}{n\hbar}} - 1\right)\delta_{w,l}
    | l\rangle  \\
    {}& = \left[1 +
     \left(e^{\frac{-itE}{n\hbar}} - 1\right)f(l)\right]
    |l\rangle. \\
\end{split}
\end{equation}
Notice that
\begin{equation}
    1 +
     \left(e^{\frac{-itE}{n\hbar}} - 1\right)f(l)
    = \left\{
       \begin{matrix}
       e^{\frac{-itE}{n\hbar}} & \text{ if } l = w \\
       1 & \text{ if } l \neq w
       \end{matrix}
      \right.  
\end{equation}
Hence the proposition follows.  \qed
\end{Proof}
\begin{Proposition} \label{Path-Prop2}
Let $|l\rangle, |m\rangle$
be basis vectors of the computational basis,
then
\begin{equation}
   \langle l |
   e^{\frac{-itE\left(|s\rangle\langle s| - I\right)}{n\hbar}}
    |m \rangle =
    e^{\frac{-itE(-1)}{n\hbar}}
    \left[\delta_{l,m} + \dfrac{1}{N}
    \left(
      e^{\frac{-itE}{n\hbar}} - 1\right)
    \right]. 
\end{equation}
\end{Proposition}
\begin{Proof}Commuting the identity matrix and then
Expanding the exponent yields
\begin{equation}
  \begin{split}
   \langle l |
   e^{\frac{-itE\left(|s\rangle\langle s| - I\right)}{n\hbar}}
    |m \rangle {}&=
    e^{\frac{-itE(-1)}{n\hbar}}
    \left[\delta_{l,m} +
    \left(e^{\frac{-itE}{n\hbar}} - 1\right)
    \langle l| s \rangle \langle s | m\rangle \right] \\
    {}& = e^{\frac{-itE(-1)}{n\hbar}}
    \left[\delta_{l,m} + \dfrac{1}{N}
    \left(
    e^{\frac{-itE}{n\hbar}} - 1\right)
   \right].  \qed
   \end{split}
\end{equation}
\end{Proof}
\begin{Proposition} 
\label{Path-Prop3}
Let $l_0 = k$ and
$l_n = j$, then The propagator in equation \ref{Propa}is
given by
\begin{equation}
   \begin{split}
    {}& K(j,k,t) =
     \lim_{n\to\infty}
    \sum_{l_1, l_2, \dots , l_{n-1}}
    \prod_{m = 0}^{n - 1}
    e^{\frac{-itE\left(f(l_m) - 1\right)}{n\hbar}}
    \left[\delta_{l_m,l_{m+1}} + 
    \frac{\left(
      e^{\frac{-itE}{n\hbar}} - 1\right)}{N}
    \right] = \\ 
  {}& \lim_{n\to\infty}
    \sum_{l_1, l_2, \dots , l_{n-1}}
     \exp{\frac{-itE\sum_{m = 0}^{n-1}\left(f(l_m) - 1\right)}{n\hbar}}
    \prod_{m = 0}^{n - 1}
    \left[\delta_{l_m,l_{m+1}} + \frac{\left(
      e^{\frac{-itE}{n\hbar}} - 1\right)}{N}
   \right]
\end{split}
\end{equation}
\end{Proposition}
\begin{Proof}
Using the Lie-Trotter produce formula, equation \ref{Propa} can be
written as
\begin{equation}
   K(j,k,t) =
  \lim_{n\to\infty}\langle j |
   \left[e^{\frac{-itE\left(|s\rangle\langle s| - I\right)}{n\hbar}}
   e^{\frac{-itE|w\rangle\langle w|}{n\hbar}} \right]^n
    |k \rangle .
\end{equation}
For each of the $n$ products, inserting the identity
\begin{equation}
  I = \sum_{l_{\alpha}} |l_{\alpha}\rangle\langle l_\alpha|,
  \quad \alpha = 1, 2, \dots n - 1 
\end{equation}
and denoting
\begin{equation}
    k \equiv l_0, \quad j \equiv l_{n} 
\end{equation}
yields
\begin{equation}
   K(j,k,t) =
  \lim_{n\to\infty}
    \sum_{l_1, l_2, \dots , l_{n-1}}
    \prod_{m = 0}^{n - 1}
    \langle l_{m+1} |
   e^{\frac{-itE\left(|s\rangle\langle s| - I\right)}{n\hbar}}
   e^{\frac{-itE|w\rangle\langle w|}{n\hbar}}
    |l_{m} \rangle .
\end{equation}
The result follows by applying propositions \ref{Path-Prop1} and 
\ref{Path-Prop2} to the last equation.  \qed
\end{Proof}

Unfortunately, in the continuous limit, a sum-over-paths interpretation
of proposition \ref{Path-Prop3}
is problematic.  The expression $\delta_{l_m, l_{m+1}}$
has the value 1 if at time $mt/n$, the state of the system
is $|l_m\rangle$ and at time $(m + 1)t/n$, the state of the system
is $|l_{m+1}\rangle$.  In the limit to continuous time,
$\delta_{l_m, l_{m+1}}$ becomes meaningless.

We proposed a nonstandard analysis formulation of proposition
\ref{Path-Prop3}
which allows us to give proposition \ref{Path-Prop3}
a sum-over-paths interpretation.
For details on nonstandard analysis, see \cite{Alb, Cut, Hur, Stroy}
 and references
within.
\begin{Theorem} 
\label{Path-Thm}
Let $\omega\in {}^*\mathbb N$
be an infinite nonstandard natural number, then
the propagator for the GFG algorithm is
given by
\begin{equation} 
  \begin{split}
  {}&K(j,k,t) =
    st\Bigg\{ \\
    {}& \sum_{l_1, l_2, \dots , l_{\omega-1}}
    \exp{
     \left[{\dfrac{-itE\sum_{m = 0}^{\omega -1}
      \left(f(l_m) - 1\right)}{\omega\hbar}}
     \right]}
    \prod_{m = 0}^{\omega - 1}
    \left[\delta_{l_m,l_{m+1}} + \dfrac{1}{N}
    \left(
      e^{\frac{-itE}{\omega\hbar}} - 1\right)
    \right]\Bigg\} .
   \end{split}
\end{equation}
\end{Theorem}
\begin{Proof} This is an application of nonstandard analysis's
formulation of limits to proposition \ref{Path-Prop3}.\qed
\end{Proof}

The interpretation of theorem 
\ref{Path-Thm} is as follows.  First, fix 
an infinite $\omega\in{}^*\mathbb N$ and time slice
$[0,t]$ into $\omega + 1$ number of steps
$T = \left\{0, t/\omega, 2t/\omega, \dots ,t\right\}$.
Notice that $\epsilon \equiv t/\omega$ is infinitesimal.
For any path
\begin{equation}
   P:T\to \left\{1, 2,\dots, N\right\}, 
\end{equation}
we associate the amplitude
\begin{equation}
  \label{Amp}
  \exp{\left[{\dfrac{-itE\sum_{m = 0}^{\omega -1}
      \left(f(P(m\epsilon)) - 1\right)}{\omega\hbar}}
     \right]}
    \prod_{m = 0}^{\omega - 1}
    \left[\delta_{P(m\epsilon),P((m+1)\epsilon)} + \dfrac{1}{N}
    \left(
      e^{\frac{-itE}{\omega\hbar}} - 1\right)
    \right] 
\end{equation}
to $P$.  The propagator for GFG 
algorithm evolution is then infinitesimally close
to the sum of the amplitudes of all possible
paths $P$.  Notice that 
for all fixed infinite
$\omega\in{}^*\mathbb N$, the amplitude of 
any path has to be of the form given in equation \ref{Amp}.

\section{\bf Semi-classical Limits\label{Semi}}
In this last section, we will investigate the
semi-classical limit behavior of the GFG and the
Grover algorithms within
the context of this paper and discuss some open problems.

One way to obtain the semi-classical limit of a quantum
system is by taking the limit $\hbar\to 0$.
In the literature of quantum computing,
$\hbar$ is usually set to the
value of 1.  Throughout this paper, we have included
$\hbar$ in all the equations.  According to section
\ref{Ana}, the GFG algorithm is obtained by
setting
$t = \dfrac{k\pi\hbar}{E}$.  Further,
the two unitary operators $U_f$ and $U_s$
were also obtained from the Lie-Trotter product
approximation by setting
$t = \dfrac{k\pi\hbar}{E}$.  Hence, in
the context of this paper, in the semi-classical
limit, both the GFG and the Grover 
algorithm must run instantaneously.  This
is because $t \to 0$ as $\hbar \to 0$.
It is not clear if this has any meaning in terms of the
standard interpretation of Grover's algorithm
in the literature.  In the literature, the unitary
operators $U_s$ and $U_f$ are time independent which
could unrealistic since quantum
mechanics is evolved by time dependent evolution
operators.  If we view decoherence due
to interactions with the environment 
as the quantum system becoming classical,  
then we might consider modeling decoherence
by $\hbar\to 0$.  If we model decoherence this way,
then the semi-classical limit behavior of
the GFG and the Grover algorithm tells us that
the algorithms must run instantaneously before
the systems become decoherent.

One could argue that semi-classical limit behaviors
are not relevant to quantum computing, but
recently there has been work done on the semi-classical
limit behaviors of Shor's algorithm, see \cite{Semi}.
The authors in \cite{Semi} proposed that it might
be worthwhile to consider building quantum
computers to run Shor's algorithm by
using semi-classical devices.  The authors showed
that in the semi-classical limit, Shor's
algorithm's success probability is not too
severely modified.  It would be interesting
to see whether techniques similar to the ones
used in this paper can be applied to Shor's
algorithm and if in the semi-classical limit,
Shor's algorithm must run instantaneously.

\end{document}